\documentclass[acmsmall,screen,nonacm]{acmart}

\usepackage[utf8]{inputenc}
\usepackage{textcomp}
\usepackage{tabularx}
\usepackage{multirow}

\usepackage{hyperref}
\usepackage[frozencache,cachedir=minted-cache]{minted}
\usepackage{xcolor} % to access the named colour LightGray
\definecolor{LightGray}{gray}{0.95}
\setminted[cpp]{
    fontsize=\tiny,%\scriptsize,
    linenos,
    breaklines,
    escapeinside=@@
}
\setminted[yaml]{
    fontsize=\tiny,%\scriptsize,
    linenos,
    breaklines,
    escapeinside=@@
}
\setminted[cmake]{
    fontsize=\scriptsize,%\scriptsize,
    breaklines,
    escapeinside=@@
}

\newcommand{\q}[1]{``#1''}
\usepackage[nolist]{acronym}

\usepackage{paralist}
\renewcommand\footnotetextcopyrightpermission[1]{}
\newcommand{\numberbubble}[1]{\raisebox{.5pt}{\textcircled{\raisebox{-.9pt} {#1}}}}

\setcopyright{acmcopyright}
\acmDOI{10.475/1234.5678}

\acmConference[ArXiv]{ArXiv}
\acmYear{2024}
\copyrightyear{2024}

\acmPrice{15.00}

\begin{document}
\begin{acronym}
\acro{simd}[SIMD]{Single Instruction Multiple Data}
\acro{mimd}[MIMD]{Multiple Instructions Multiple Data}
\acro{sise}[SISE]{SIMD instruction set extension}
\acroplural{sise}[SISEs]{SIMD instruction set extensions}
\acro{sfinae}[SFINAE]{Substitution failure is not an error}
\acro{sru}[SRU]{SISE representation unit}
\acroplural{sru}[SRUs]{SISE representation units}

\end{acronym}
\title{Designing and Implementing a Generator Framework for a SIMD Abstraction Library}
%\title{SIMD Abstraction Library - Design and Implementation of a Generator Framework}
\author{Johannes Pietrzyk}
\affiliation{%
  \institution{TU Dresden, Database Systems Group}
  \city{Dresden}
  \state{Germany}
}
\email{firstname.lastname@tu-dresden.de}

\author{Alexander Krause}
\affiliation{%
  \institution{TU Dresden, Database Systems Group}
  \city{Dresden}
  \state{Germany}
}

\author{Dirk Habich}
\affiliation{%
  \institution{TU Dresden, Database Systems Group}
  \city{Dresden}
  \state{Germany}
}

\author{Wolfgang Lehner}
\affiliation{%
  \institution{TU Dresden, Database Systems Group}
  \city{Dresden}
  \state{Germany}
}

\renewcommand{\shortauthors}{J. Pietrzyk et al.}

\begin{abstract}
The \emph{Single Instruction Multiple Data (SIMD)} parallel paradigm is a well-established and heavily-used hardware-driven technique to increase the single-thread performance in different system domains such as database or machine learning. 
Depending on the hardware vendor and the specific processor generation/version, SIMD capabilities come in different flavors concerning the register size and the supported SIMD instructions. 
Due to this heterogeneity and the lack of standardized calling conventions, building high-performance and portable systems is a challenging task. 
To address this challenge, academia and industry have invested a remarkable effort into creating SIMD abstraction libraries that provide unified access to different SIMD hardware capabilities. 
However, those one-size-fits-all library approaches are inherently complex, which hampers maintainability and extensibility. 
Furthermore, they assume similar SIMD hardware designs, which may be invalidated through ARM SVE's emergence. 
Additionally, while existing SIMD abstraction libraries do a great job of hiding away the specifics of the underlying hardware, their lack of expressiveness impedes crucial algorithm design decisions for system developers. 
To overcome these limitations, we present TSLGen, a novel end-to-end framework approach for generating an SIMD abstraction library in this paper.
We have implemented our \emph{TSLGen} framework and used our generated \emph{Template SIMD Library (TSL)} to program various system components from different domains. 
As we will show, the programming effort is comparable to existing libraries, and we achieve the same performance results. 
However, our framework is easy to maintain and to extend, which simultaneously supports disruptive changes to the interface by design and exposes valuable insights for assessing provided functionality.
\end{abstract}

\maketitle
\sloppy

\section{Introduction}
\label{sec:Introduction}

\emph{\acf{simd}} is a well-established parallelism concept that is characterized by the fact that the \emph{same operation} is simultaneously applied on \emph{multiple data elements} within a single instruction~\cite{DBLP:series/synthesis/2015Hughes}. 
This state-of-the-art paradigm plays a crucial role in increasing the single-thread performance in data-centric as well as compute-heavy domains such as Database Systems (DB)~\cite{DBLP:conf/sigmod/PolychroniouRR15,DBLP:journals/vldb/PolychroniouR20,DBLP:conf/systor/ZarubinDKHL21}, Machine Learning Systems (ML)~\cite{DBLP:conf/icdm/BohmP15,DBLP:conf/iscc/CarneiroSN21,DBLP:journals/eaai/DiasAM04} or High-Performance Computing (HPC)~\cite{DBLP:conf/sc/ChhuganiKSPDSS12,DBLP:journals/concurrency/0001FLM21,DBLP:journals/superfri/KomatsuOFFIMSK21}. 
Modern CPUs provide evolving support for SIMD capabilities using vendor-specific SIMD instruction set extensions (SISE). 
Those extensions usually include arithmetic and Boolean operators, logical and arithmetic shifts, and data type conversions, including specific SIMD instructions to load data from main memory into SIMD registers and write it back. 
Furthermore, the availability of SIMD instructions for a particular functionality or data type varies depending on ARM, Intel, and other vendors.
This leads to a highly heterogeneous SIMD landscape. 

As porting code to different hardware is a laborious and costly task, industry and research have invested considerable effort and time into designing and developing SIMD abstraction libraries to first and foremost address the portability challenge of highly optimized SIMD code.
Examples are -- making no claim to be exhaustive -- XSIMD~\cite{xsimd}, Google Highway~\cite{highway},  TVL~\cite{TVLPaper}, or the Generic SIMD Library~\cite{DBLP:conf/ppopp/WangWTSM14}.
Those libraries are usually implemented in C/C++ for performance reasons, and they leverage C++ class templates to represent a SIMD hardware register. 
They all have in common that they hide the concrete implementation (or workaround) behind a dedicated function call. Furthermore, the templates enable compile-time deduction and code generation with zero overhead for the runtime.
Leveraging such libraries allows writing easily exchangeable code to execute the same algorithm with different SIMD SISEs. 
However, as those libraries are software systems on their own, they must be implemented and extended whenever additional functionalities are required or new hardware emerges. 
Consequently, the burden of complexity now manifests in designing and implementing such a library. 
The existing solutions are all one-size-fits-all hand-crafted libraries, meaning that every supported hardware is defined inside the library. 
However, building, maintaining, and extending of such a SIMD abstraction library is non-trivial and requires, e.g., carefully placed \#\textsc{ifdef} statements to ensure seamless portability between platforms.
This leads to significant drawbacks like high code redundancy and poor readability through complex preprocessor conditionals. 
Furthermore, evolving hardware frequently entails disruptive changes in processing paradigms. 
To give two examples, first, with the advent of Intel AVX512~\cite{avx512}, a new SIMD-register mask type alongside masked operations was introduced, changing not only the return type of specific functions but also introducing a whole new category compared to SSE and AVX(2). 
Second, when ARM introduced its Scalable Vector Extension (SVE)~\cite{ARMSVE}, the former assumption of a given SIMD register size could no longer be sustained. 
Those fundamental changes frequently entail the need to revise past design decisions. 
This often incurs a high refactoring effort concerning the API and the underlying implementations.

\textbf{Our Contribution and Outline:}
Code generation is a well-established technique to cope with the drawbacks of one-size-fits-all libraries. 
In this paper, we present a novel design and an implementation of a framework called \emph{TSLGen} for generating a SIMD abstraction library.
Based on that, we make the following contributions in this paper: 
\begin{compactenum}
\item In Section~\ref{sec:Background}, we present related work in the context of SIMD abstraction libraries in more detail. In particular, we introduce requirements for such libraries and discuss the manual approach of existing libraries to our generator approach.
\item Afterwards, we introduce our novel developed framework \emph{TSLGen} in Section~\ref{sec:TSLGen}. We start with a schematic overview before describing the individual components of our framework in more detail. As we will show, our overall approach is based on the complete generation of a SIMD abstraction library.
This makes the library much easier to maintain and extend, and it even allows for cherry-picking of only a few specific parts to be generated. Thus, we can generate the complete library or only a slim one on a per-use-case basis whenever necessary.
\item Then, we introduce advanced features extending the functionalities of \emph{TSLGen} in Section~\ref{subsec:tsl_extension}. In particular, we discuss extensions for test generation, build environments and tool support. 
\item In Section~\ref{sec:Evaluation}, we present selected evaluation results based on use cases from different perspectives\footnote{The implementation of our \emph{TSLGen} framework, including several use-case implementations, will be made available in case of acceptance.}. In particular, we show that we can seamlessly integrate Intel FPGA accelerator cards as flexible SIMD processing units in addition to classical SIMD CPU extensions. 
This facilitates the usage of such accelerator cards since no specific code base needs to be implemented anymore.  
This is a big step forward in executing efficient arbitrary CPU-SIMD code on FPGA.  
\item Subsequently, we discuss additional opportunities of our generator approach in Section~\ref{sec:Discussion}.
\end{compactenum}
Finally, we discuss related work in a broader context in Section~\ref{sec:RelatedWork} and close the paper with a summary in Section~\ref{sec:Conclusion}.

\section{Background}
\label{sec:Background}
The major motivation behind using \ac{simd} in systems is to increase single-thread performance~\cite{DBLP:journals/vldb/PolychroniouR20,TVLPaper}. 
For accessing \ac{simd} capabilities provided by the hardware, different approaches exist: (i) using auto-vectorization features of modern compilers, (ii) using assembly, and (iii) using hardware-specific intrinsics. 
All of the techniques mentioned above have their specific advantages and drawbacks. 
On the one hand, auto-vectorization can be used in a hardware-agnostic manner, thus significantly reducing the necessary effort of porting existing code. 
On the other hand, specific patterns are sometimes not recognized as simdifiable by the compiler, leading to the need for continuous validation of the compilation result. 
In contrast, assembly enables the precise use of existing hardware functionalities. 
However, this freedom comes at the price of platform dependency on the one hand and relatively high complexity (e.g., manual register allocation) on the other. 
While auto-vectorization and assembly are at opposite ends on the abstraction spectrum, using intrinsics can be located somewhere in between since they offer a fine-grained, explicit method of expressing the desired behavior while abstracting away the most low-level details. 
Yet, these intrinsics are primarily designed to map the underlying \ac{sise} directly.
On the one hand, this enables the best performance for a specific \ac{simd} platform. 
However, this also leads to the lack of portability across different \ac{simd} platforms, on the other hand.
Given the benefits of intrinsics, their use is a de-facto standard in various domains~\cite{DBLP:conf/icdm/BohmP15,TVLPaper,DBLP:conf/ppopp/WangWTSM14}.   
To explore different \ac{simd} platforms, system developers have to implement multiple versions of the same system component, one for each target platform, as shown for example for the comparison operator in Figure~\ref{lst:cmpeq_sse_neon}.
Here, a \ac{simd} register comparison using Intel's SSE and ARM's NEON producing an integral mask is depicted. 
Both implementations produce a \ac{simd} register, where all bits are set to 1 within an entry (henceforth denoted as lane) if the corresponding input lanes are equal. Otherwise, all bits are set to 0. 
This \ac{simd} register is used to form a mask, where a 1-bit at the n-th position indicates that the n-th values from both input registers were binary identical.
As shown, the same functionality is implemented differently which significantly hinders portability. 

\begin{figure}
    \begin{minipage}[t]{.49\linewidth}
        \centering 
        %\lstinputlisting[frame=off,language=C++]{#2}
\begin{minted}{cpp}
uint8_t equal(__m128i v0, __m128i v1) {
  __m128i m   = _mm_cmpeq_epi32(v0, v1);
  __m128  tmp = _mm_castsi128_ps(m);
  return _mm_movemask_ps(tmp);
}
\end{minted}
        \small (a) Intel SSE implementation\\
    \end{minipage}
    \hfill
    \begin{minipage}[t]{.49\linewidth}
        \centering 
        %\lstinputlisting[frame=off,language=C++]{#4}
\begin{minted}{cpp}
uint8_t equal(uint32x4_t v0, uint32x4_t v1) {
  uint32x4_t m = vceqq_u32(v0, v1);
  uint32_t buf[4];
  vst1q_u32(&buf, m);
  return 
    (1&buf[0]) | (2&buf[1]) | (4&buf[2]) | (8&buf[3]);
}
\end{minted}
        \small (b) ARM Neon implementation\\
    \end{minipage}
    \caption{Illustration of \ac{simd} heterogeneity.}
    \label{lst:cmpeq_sse_neon}
    \vspace{-0.4cm}
\end{figure}

\begin{figure}[b]
    \vspace{-0.4cm}
    \centering
    \includegraphics[width=0.7\linewidth]{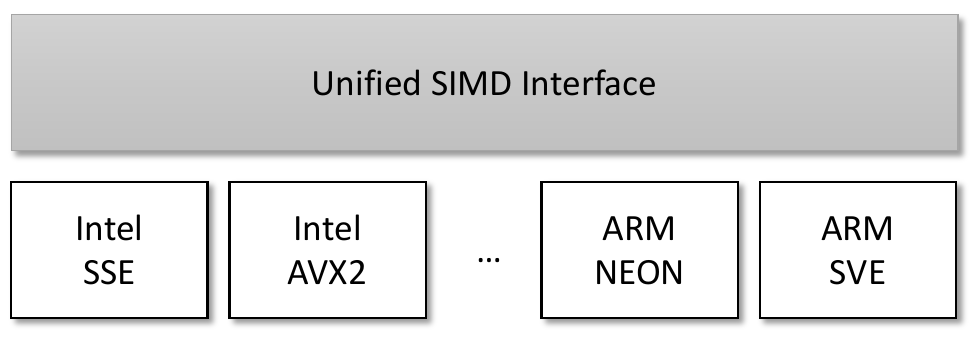}
    \caption{Common \ac{simd} abstraction library architecture according to ~\cite{TVLPaper} and~\cite{DBLP:conf/ppopp/WangWTSM14}.}
    \label{fig:CommonApproach}
    \vspace{-0.4cm}
\end{figure}

To overcome this portability issue, many \ac{simd} abstraction libraries have been developed in recent years~\cite{DBLP:conf/IEEEpact/EsterieGFLR12,highway,TVLPaper,xsimd,DBLP:conf/ppopp/WangWTSM14}.
All these libraries share the same key concept as illustrated in Figure~\ref{fig:CommonApproach} by defining an abstraction layer -- unified \ac{simd} interface -- to hide platform-specific \ac{simd} register types and \ac{simd} intrinsics as well as provide abstract \ac{simd} registers and common -- abstract -- \ac{simd} operations over them. 
Moreover, all these libraries are implemented in C/C++ to avoid losing performance. 
Additionally, these libraries mainly rely on template metaprogramming in C/C++, and, e.g., abstract \ac{simd} operations are provided as function templates. 
The corresponding mapping is conducted through function template specializations. 
These function template specializations have to be implemented within these libraries, whereby the implementation depends on the available functionality of the \ac{simd} platform. 
However, this is independent of the application and must be done \emph{only once} by a domain expert for a specific \ac{simd} extension.

\subsection{Applicability}
Generally, using these \ac{simd} abstraction libraries can be very advantageous in terms of maintainability, and portability of time-sensitive parts of the system component code. 
Nevertheless, without any claim to comprehensiveness, such libraries are complex entities, designed in an up-to-date way and implemented manually (hand-written). 
Consequently, these libraries themselves struggle to ensure maintainability and uphold extensibility through increasing code complexity. 
To illustrate this, we selected a code snippet from an industrial-scoped library -- Google Highway (HWY)\footnote{\url{https://github.com/google/highway/blob/5c296e8e2a2648f1296327e11719eadc6c9438b6/hwy/ops/x86_256-inl.h} (accessed: November, 16th 2023)} (see Listing~\ref{lst:msb_extract_ghighway} a). 
While this is just an excerpt, we found it to be a profound example for investigating design decisions common to most \ac{simd} abstraction libraries. 
The depicted code is a C++ template specialization for a generalization of the last part of the previously introduced algorithm from Listing~\ref{lst:cmpeq_sse_neon}, namely the extraction of the most significant bit (MSB) of every data element within a given \ac{simd} register. 
The presented specialization is for 256-bit wide Intel AVX \ac{simd} registers to process 2-byte wide data elements.
To use this function, it is sufficient to call \emph{BitsFromMask<uint16\_t>(mask)}, while the compiler will generate the appropriate template specialization. 
Hence, if the algorithm is executed using another \ac{sise} or platform, it is sufficient to rebuild the code with the specific header files containing preprocessor macros and template instantiations, substantially reducing the manual effort of porting existing code. 

\begin{figure}
    \begin{minipage}[t]{.49\linewidth}
        \centering 
\begin{minted}{cpp}
template <typename T, HWY_IF_T_SIZE(T, 2)>
HWY_INLINE uint64_t BitsFromMask(const Mask256<T> mask) {
#if !defined(HWY_DISABLE_BMI2_FMA) && !defined(HWY_DISABLE_PEXT_ON_AVX2)
  const Full256<T> d;
  const Full256<uint8_t> d8;
  const Mask256<uint8_t> mask8 = MaskFromVec(BitCast(d8, VecFromMask(d, mask)));
  const uint64_t sign_bits8 = BitsFromMask(mask8);
  // Skip the bits from the lower byte of each u16 (better not to use the
  // same packs_epi16 as SSE4, because that requires an extra swizzle here).
  return _pext_u32(static_cast<uint32_t>(sign_bits8), 0xAAAAAAAAu);
#else
  // Slow workaround for when BMI2 is disabled
  // Remove useless lower half of each u16 while preserving the sign bit.
  // Bytes [0, 8) and [16, 24) have the same sign bits as the input lanes.
  const auto sign_bits = _mm256_packs_epi16(mask.raw, _mm256_setzero_si256());
  // Move odd qwords (value zero) to top so they don't affect the mask value.
  const auto compressed = _mm256_castsi256_si128(
      _mm256_permute4x64_epi64(sign_bits, _MM_SHUFFLE(3, 1, 2, 0)));
  return static_cast<unsigned>(_mm_movemask_epi8(compressed));
#endif  // HWY_ARCH_X86_64
}
\end{minted}
        \small (a) Latest version from 2023\\
    \end{minipage}
    \hfill
    \begin{minipage}[t]{.49\linewidth}
        \centering
        %\lstinputlisting[frame=off,language=C++]{#4}
\begin{minted}{cpp}
template <typename T, HWY_IF_LANE_SIZE(T, 2)>
HWY_INLINE uint64_t BitsFromMask(const Mask256<T> mask) {
#if HWY_ARCH_X86_64
  const Full256<T> d;
  const Full256<uint8_t> d8;
  const Mask256<uint8_t> mask8 = MaskFromVec(BitCast(d8, VecFromMask(d, mask)));
  const uint64_t sign_bits8 = BitsFromMask(mask8);
  // Skip the bits from the lower byte of each u16 (better not to use the
  // same packs_epi16 as SSE4, because that requires an extra swizzle here).
  return _pext_u64(sign_bits8, 0xAAAAAAAAull);
#else
  // Slow workaround for 32-bit builds, which lack _pext_u64.
  // Remove useless lower half of each u16 while preserving the sign bit.
  // Bytes [0, 8) and [16, 24) have the same sign bits as the input lanes.
  const auto sign_bits = _mm256_packs_epi16(mask.raw, _mm256_setzero_si256());
  // Move odd qwords (value zero) to top so they don't affect the mask value.
  const auto compressed =
      _mm256_permute4x64_epi64(sign_bits, _MM_SHUFFLE(3, 1, 2, 0));
  return static_cast<unsigned>(_mm256_movemask_epi8(compressed));
#endif  // HWY_ARCH_X86_64
}
\end{minted}
        \small (b) Version from 2022\\
    \end{minipage}
    \caption{Evolution of MSB-extract, taken from Google's \ac{simd}-Library Highway (HWY). The function extracts the most significant bit of every value within a \ac{simd} register, and builds an integral value from those bits.}
    \label{lst:msb_extract_ghighway}
\end{figure}

\subsection{Evolving API}
Extending an existing library should require reasonable manual effort to generate real business value. 
While the presented code snipped in Figure~\ref{lst:msb_extract_ghighway} is just an exemplary case, it speaks volumes about the general design. 
All the complexity from our motivating Listing~\ref{lst:cmpeq_sse_neon} is shifted into the template specializations. 
Consequently, for developing such a library, the manual effort stays the same compared to the manual porting of existing code. 
In addition, since such libraries try to offer a \q{one-size-fits-all} solution, the resulting code contains hardwired dependencies, e.g., the preprocessor macros enabling a specific code path, depending on a compile-time flag. 
Thereby, the compiler decides at compile time what fractions of the code are used to generate the binary output. 
To achieve this behavior, the most common approach relies on using preprocessor macros (see Figure~\ref{lst:msb_extract_ghighway}). 
%While this behavior can be achieved by multiple approaches, e.g., template substitution using \emph{SFINAE\footnote{Substitution Failure is not an error}} or with the introduction of C++-17, \emph{constexpr if}, HWY uses preprocessor macros (see Figure~\ref{lst:msb_extract_ghighway}). 
Such constructions hamper the general readability since they increase the amount of dead code. 
However, what is even worse: extending the library to integrate future hardware will entail a reiterated examination of mutual exclusions and possibly even revising a substantial portion of the existing library. 

Another general drawback throughout hand-written libraries comprises structural redundancy. 
Every template specialization has to follow the same schema. 
Consequently, if the template or function signature changes, all template specializations must also be modified. 
The effects can be seen by comparing an older\footnote{\url{https://github.com/google/highway/blob/7a5ae370e83fbfea433a90cbe4ab2c88dfb59126/hwy/ops/x86_256-inl.h} (accessed: October, 20th 2022)} version of the MSB-extract with the latest one (see Figure ~\ref{lst:msb_extract_ghighway}). 
Two substantial things changed over time: (i) the preprocessor macros and (ii) the \ac{sfinae} condition of the template specialization. 
Both changes require a holistic adaption of the library, which entails a substantial manual effort. 
Another existing, yet not very apparent, problem may result from the function's return type in the future. 
In the depicted example, the function's return type is an unsigned 64-bit integer. 
The return value can hold up to 64 bits, sufficient for 512-bit \ac{simd} registers with a minimal data type size of 8 bits (i.e., 64 lanes). 
However, with the advent of ARM's SVE extension~\cite{ARMSVE}, \ac{simd} registers with a size of up to $2048$ bits are possible -- in the more extreme case, considering the accelerator card SX Aurora TSUBASA~\cite{DBLP:conf/sc/KomatsuMIWMYA0K18}, registers consist of  $16384$ bits. 
If such extensions are included in the library, the resulting mask types will need much more space than 64-bit ($256$ bit for SVE, $2048$ bit for SX Aurora TSUBASA, respectively). 
Consequently, the function signature and all the template specializations must be revised.

\section{TSLGen - Generative Framework}
\label{sec:TSLGen}

\begin{figure}
    \begin{minipage}[t]{.49\linewidth}
        \centering 
\begin{minted}{cpp}
struct {{ SRU_name }} {
  template<
    TSLArithmetic BaseType, 
    std::size_t RegSize = {{ SRU_default_size }}
  >
    struct types {
      using size_b = 
        std::integral_constant<std::size_t, RegSize>;
      using register_t {{ SRU_reg_attr }} = 
        {{ SRU_reg_t }};
      using mask_t     {{ SRU_mask_attr }} = 
        {{ SRU_mask_t }};
      using imask_t    {{ SRU_imask_attr }} = 
        {{ SRU_integral_mask_t }};
    };
};
\end{minted}
        \small (a) \ac{sru} struct code template example.\\
    \end{minipage}
    \hfill
    \begin{minipage}[t]{.49\linewidth}
        \centering 
        %\lstinputlisting[frame=off,language=C++]{#4}
\begin{minted}{yaml}
---
vendor: "intel"
extension_name: "sse"
SRU_default_size: 128
SRU_reg_attr: |-
  __attribute__((
    __vector_size__(RegSize/8),__may_alias__,
    __aligned__(RegSize/8)
  ))
SRU_reg_t: "BaseType"
SRU_mask_attr: ""
SRU_mask_t: "register_t"
SRU_imask_attr: ""
SRU_integral_mask_t: |
  std::conditional_t<sizeof(BaseType)==1, uint16_t, uint8_t>
\end{minted}
        \small (b) User-provided input for an SSE-\ac{sru}\\
        ~\\
    \end{minipage}
    \caption{Example for separation of API-specific template from specific user provided data model.}
    \label{lst:extension_template}
\end{figure}

To cope with the challenges entailed by hand-written \ac{simd} libraries, we propose replacing the manual approach with a well-defined generator approach.
%To present our framework, we start with a comprehensive introduction to the used terminology and the general core concepts of code-generator frameworks. 
%Afterwards, we introduce our framework. We start with a schematic overview before describing the individual components in more detail. 
%\subsection{Terminology}
As we already described, there are multiple vendor-specific \ac{sise}. 
Within our generator framework, we abstract those \acp{sise} using C++ structs called \ac{sru}. 
On the one hand, every \ac{sise} provides a variety of \ac{simd} functionalities called intrinsics operating on \ac{simd} registers or masks. 
On the other hand, \ac{simd} abstraction libraries offer a unified access model to those intrinsics. 
Typically, this is realized through functions. 
We call these functions primitives to differentiate them from actual intrinsics. 
%This is done to highlight that a primitive may directly map to an intrinsic, but if the given \ac{sise} does not provide the required intrinsic, implementing the primitive will include either scalar code or a workaround using other intrinsics.  
A primitive can either map directly to a single intrinsic or to a combination of intrinsics for more complex functionality.
If a \ac{sise} does not provide a required intrinsic, a workaround using other intrinsics or scalar code has to be employed to provide the same functionality.

\subsection{General Concepts}
Without loss of generality, code generation frameworks typically consist of three components: (i) \emph{code-specific skeletons} (aka. \emph{code templates}) are used for the generation process, (ii) \emph{target-specific data}, which is used to populate the code templates, and (iii), general functionality to load and process the formerly mentioned components. 
More specifically, \emph{code templates} contain all the static information relevant to the generated code. 
Since we utilize code-generation features for creating a C++ template \ac{simd} abstraction library, we require the aforementioned code templates for at least three cases: (i) defining an \ac{sru}, (ii) declaring, and (iii) defining a primitive. 
Figure~\ref{lst:extension_template} (a) shows an exemplary template for an \ac{sru} extension. 
Without going into detail, the skeleton consists of specific C++ code (e.g., lines 2-3, 5-7) and placeholders encapsulated in double curly brackets. 
Those placeholders are populated using \emph{target-specific data} (see Figure~\ref{lst:extension_template} (b)).
Without restriction of generality, the data used to populate code templates often are dictionaries where the relevant placeholder names are keys, and the data that should be used for substitution is the value.

\subsection{Framework Description}
\begin{figure*}
    \includegraphics[width=0.5\linewidth]{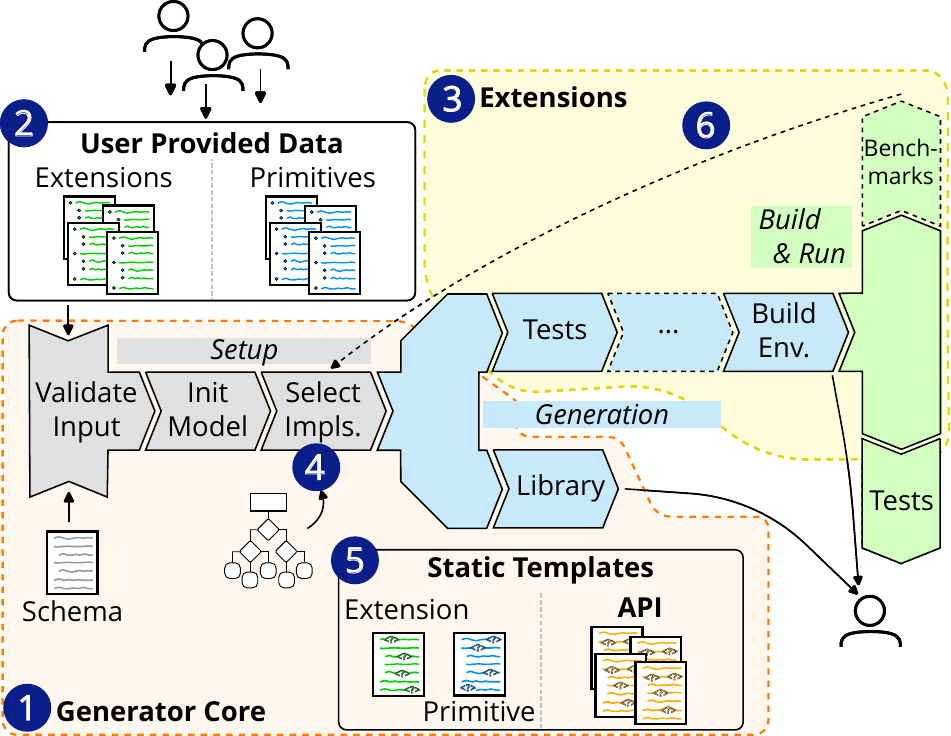}
    \caption{Architecture overview of our \emph{TSLGen} framework.}
    \label{fig:tslgen_overview}
    \vspace{-0.3cm}
\end{figure*}

Figure~\ref{fig:tslgen_overview} provides a schematic overview of \emph{TSLGen}, and as illustrated, our framework consists of three main components. 
First, the \emph{TSLGen Core} is the framework's eponymous component and the central element of the library generation process ~\numberbubble{1} (numbers in circles are
referring to the numbers in the blue circles in Figure~\ref{fig:tslgen_overview}). 
Second, external data, created and maintained by library developers or users (from now on, we will subsume those under the term users), serves as input for the generator~\numberbubble{5}. 
Finally, as the core consists of pipelined stages, a designated port exists for plugging in multiple extensions, e.g., to ensure the quality of service (e.g., test case generation) and support quality of life features~\numberbubble{7}. 
In the following, we will describe each component in more detail.

%%%%%%%%%%%%%%%%%%%%%%%%%
%% TSLGen Core
%%%%%%%%%%%%%%%%%%%%%%%%%
\textbf{\numberbubble{1} TSLGen Core}
We designed our generator core as a pipeline consisting of multiple generator pipeline operators (GPO), where every GPO depends on the result of the previous one. 
That way, the GPOs remain exchangeable, and the pipeline can be altered in its behavior by changing an operator or expanded by adding further operators (see Section~\ref{subsec:tsl_extension}). 
The fundamental core pipeline consists of four GPOs~\numberbubble{1}. 
The very first GPO validates the input provided to the generator. 
While this step may be omitted, it can be very beneficial when searching for errors within the input and enriching the provided user data (see description of~\numberbubble{5}). 
With GPOs~\numberbubble{1}, every code template is loaded once into the framework and subsequently validated.
The resulting data is then either passed on to the next GPO or errors are prompted to the user.
%After the input is validated, every user-defined \ac{sru} and primitives with all associated \ac{simd}-specific implementations is loaded once forming a complete representation of the library data model. 
The next GPO in the pipeline selects the \acp{sru} and primitives from the data model, which are relevant to the target hardware~ \numberbubble{2}. 
The necessary information about the hardware is either provided as user input or can be directly queried from the operating system by the GPO itself. 
By only generating relevant \acp{sru} and primitive implementations, we can substantially decrease the complexity of the library, supporting readability and eventually decreasing the compile time. 
In the simplest scenario, the user input contains only a single implementation $p_0$ for a given primitive $p$ and a specific \ac{sru}. 
However, as described above, there may exist multiple implementations $p_i$ for specific subsets of the target \acp{sise} (or even different algorithmic flavors, that will be discussed in more detail in Section~\ref{subsec:further_extensions}), which would be well-formed and valid on the underlying hardware. 
However, generating code for all variants would yield multiple definitions of the same function, which is usually prohibited by most languages.
Hence, the GPO has to decide which of the existing implementations is used for library generation. 
There are numerous well-known potential strategies for the selection process, like a simple pick-first scheme over a heuristic model or even a syntactic analyzer, to name just two. 
For our \emph{TSLGen}, we implemented a heuristic model, which finds the highest match between the required hardware capabilities of the user given implementation and the actually available hardware features. 
The underlying idea is that if an implementation uses more hardware-provided functionalities, the implementation utilizes the given hardware circuits more optimal or, in other words, is more specialized against the underlying hardware. 
If multiple variants with the same similarity score exist, the implementations are sorted ascending by the number of lines of code, and the first (i.e. shortest) implementation is chosen. 
While this strategy may not lead to an optimal selection result, our framework enables an uncomplicated exchange of the strategy by design.

The final GPO in our \emph{TSLGen}-pipeline generates the library code~\numberbubble{3}. 
The generation-operator consists of two steps. 
First, all available \acp{sru} are created as classes. 
Second, for every primitive, which has a given implementation for the target hardware, a helper C++ class template, and a function is generated. 
The function will be exposed as a primitive to the library user and forwards the input to the helper class. 
The helper class will have C++ template specializations for every existing implementation, depending on the \ac{sru} and the datatype. 
The distinction between the function template and the to-be-specialized helper class has two significant benefits. 
First, a class can be used as a functor, which can be helpful if algorithms from the C++ standard template library (STL) are used. 
Second, contrary to functions, C++ classes can be partially specialized, which can reduce the number of specializations significantly and help mitigate code complexity. 

Instead of implementing a monolithic generation operator with complex rules and dependencies, we propose to separate the concern of generation from the actual rules determining how the generation should be carried out. 
This can be realized with modern template engines, e.g., Jinja2~\cite{jinja2}. 
Thus, our developed framework \emph{TSLGen} is implemented in Python generating a C++ \ac{simd} abstraction library.  
Those template engines take a template and a data model (UPD) as input and produce the result document. 
Thus, the template contains placeholders that are substituted with the data from the given model at generation time (cf. Figure~\ref{lst:extension_template}~(a)). 
Even basic control-flow logic can be embedded in the generator templates, depending on the information of the given UPD. 
Consequently, the business logic resides within those templates as a single point of truth and can be adapted easily. 
Using templates for code generation offers a high degree of freedom since potential future changes to the template do not require modification within the corresponding GPOs. 
Furthermore, while we currently focus on C++-library code generation, our framework can potentially support other programming languages by changing the templates accordingly and update the UPD.

%%%%%%%%%%%
%% Use provided Data
%%%%%%%%%%%
\textbf{\numberbubble{5} User provided Data (UPD)}
For the data model, Jinja2 expects a dictionary containing keys that map to the parameter names equal to the placeholders of the template. 
The values of those keys will be used for substituting the placeholders. 
Consequently, all values have to be present for rendering the template. 
To ensure that the UPD satisfies the implicitly existing data model, we manually inferred a schema from the templates using a bottom-up approach~\numberbubble{6}\footnote{The inference could be also done automatically to reduce error susceptibility}. 

\textbf{\numberbubble{6} Schema Description}
The schema consists of several entries. 
Every entry has a name and an expected fundamental (e.g., string or a list of strings) or composed type. 
A composed type consists of multiple entries. 
To improve the unambiguity of the schematized data model, we distinguish between two types of entries within a composed type. 
First, mandatory entries must be specified within the input. 
Second, optional entries may or may not be specified. 
%Nevertheless, a default value is defined for every optional entry of the schema. 
Despite being optional, these values are necessary for the validation phase to enrich the provided data, hence a default value is defined for every optional entry. 
Thereby, we can significantly decrease the number of necessary entries for the UDP. 
We also allow arbitrary additional fields beyond the ones specified by the schema, which can be instantiated and employed by the user.

\begin{figure}
    \begin{minipage}[t]{.49\linewidth}
        \centering 
\begin{minted}{yaml}
---
primitive_name: "to_integral"
parameters:
 - ctype: "const typename SimdT::mask_type"
   name: "mask"
returns:
 ctype: "typename SimdT::imask_type"
#...#
definitions:
#...#
 - target_extension: "sse"
   ctype: ["int16_t", "uint16_t"]
   lscpu_flags: ["sse", "sse2", "bmi2"]
   implementation: |
     return 
       _pext_u32(_mm_movemask_epi8(mask), 0x5555);
 - target_extension: "sse"
   ctype: ["int16_t", "uint16_t"]
   lscpu_flags: ["sse", "sse2"]
   implementation: |
     return
       _mm_movemask_epi8(
         _mm_packs_epi16(mask, _mm_setzero_si128()));
#...#
...
\end{minted}
        \small (a) User-provided input for MSB-extract.\\
    \end{minipage}
    \hfill
    \begin{minipage}[t]{.49\linewidth}
        \centering 
        %\lstinputlisting[frame=off,language=C++]{#4}
\begin{minted}{cpp}
namespace details {
  template<> struct to_integral<simd<uint16_t, sse>> {
    using SimdT = simd<uint16_t, sse>;
    [[nodiscard]] SIMDLIB_FORCE_INLINE static 
    typename SimdT::imask_type apply(
      const typename SimdT::mask_type mask
    ) {
      /** 
       * If hardware supports bmi2, 
       * the following code is generated.
       **/
      return 
        _pext_u32(_mm_movemask_epi8(mask), 0x5555);
      /** 
       * Otherwise, the following code 
       * would be generated.
       * return 
       *   _mm_movemask_epi8(
       *     _mm_packs_epi16(mask, _mm_setzero_si128())
       *   );
       **/
    }
  };
}
\end{minted}
        \small (b) Generated Template instantiation for SSE, uint16\_t\\
    \end{minipage}
    \caption{Example of user provided data and the resulting generated C++-code.}
    \label{lst:primitive}
\end{figure}

\textbf{\numberbubble{5} User provided Data (UPD) - Input Description}
Efficient schema-validation requires \mbox{(semi-)structured} data. 
While there exist various data formats, we decided to use YAML~\cite{yaml}. 
YAML is a key-value format that clusters associated data structurally by using indentation. 
Doing so increases the readability compared to other \mbox{(semi-)structured} formats like XML or JSON. 
In addition, YAML is widely supported and provides all required data types natively. 
Unfortunately, YAML does not come with a schema-DSL, so we had to implement a validation-GPO ourselves. 
An example of such an UDP file is depicted in Figure~\ref{lst:primitive} (a). 
The presented excerpt contains the data for a primitive with the name \emph{to\_integral}. 
This primitive is the equivalent to the function from Figure~\ref{lst:msb_extract_ghighway}. 
A single YAML document, enclosed by three dashes at the beginning and three dots at the end, exists for every primitive. 
Mandatory entries of every primitive-YAML-document are the \emph{primitive name}, a list of \emph{parameters}, a \emph{return} type and a list of \emph{definitions}. 
While the name is just a string, all other fields are composed types. 
Thereby, more complex dependencies can be expressed, e.g., a parameter consists of a name and a type but can also have associated attributes. 

The \ac{sru}-specific definitions must be provided as a list of composed types under the key \emph{definitions}. 
Every definition must specify an \ac{sru}, a list of associated C-types, a list of required \acp{sise}, and an implementation. 
In the provided example, there are two definitions for the \ac{sru} called \emph{sse} (which is the abstraction for the Intel SSE \ac{sise}) operating on 16-bit wide integer data (see Figure~\ref{lst:primitive} (a)). 
Both definitions differ in the implementation parts, either using the \emph{\_pext\_u64} intrinsic or the \emph{\_mm\_pack\_epi16} intrinsic. 
The first one is only available on hardware, which supports Intel's bit manipulation instruction sets (BMI2). 
Consequently, on x86 hardware supporting at least SSE and SSE2, the first implementation would prevail over the latter in the generated library if the hardware provides BMI2, or vice versa otherwise. 
In addition, we introduced an optional \emph{is\_native} boolean field (defaulted to True), indicating whether the implementation directly maps to the hardware provided intrinsic without employing artificial workarounds to mimic expected behavior. 
This can be very helpful for users of the generated library to reason about the performance of an algorithm, or for tailoring an implementation for a specific \ac{sise}. 
However, both implementations are not native, so a compiler warning will be generated when building the library. 

The majority of \ac{simd}-intrinsics follow a distinctive naming scheme. 
As shown in Figure~\ref {lst:primitive} (a), the intrinsics from Intel SSE are prefixed with \emph{\_mm\_} and suffixed with the compute-granularity, where \emph{\_epi16} indicates that the instruction considers 16-bit integer values within a \ac{simd} register as input. 
The same applies for other granularities like 8-bit integer values (\emph{\_epi8}) or complete 128-bit \ac{simd} registers (\emph{\_si128}).  
AVX(2) introduces the prefix \emph{\_mm256\_} and AVX512 \emph{\_mm512\_} respectively. 
ARM's \ac{sise} Neon (aka. Advanced SIMD or ASIMD) follows a similar yet more fine-grained naming schema, encoding the value type and size within the function name. %providing a distinction between unsigned and signed value operations for every supported bit width. 

We can leverage the described structural similarity of intrinsics to reduce the number of lines of code that must be written to generate all implementations further. 
To do so, we split our generation GPO into two stages.
The first stage creates a Jinja2 template from the implementation of every definition and uses the associated user-provided primitive and extension data as data model for rendering. 
Consequently, we can use Jinja2 syntax within the implementation field of the primitive definition. 
Thereby, we can write a single definition for all value types at once. 
While this can be seen as an extension to our \emph{TSLGen} pipeline, we found this simplification particularly valuable; that is why we hard-wired this feature into the generation GPO.

%%%%%%%%
%% Generator Extensions
%%%%%%%%%
\section{Advanced Features}
\label{subsec:tsl_extension}
%Takeaways:
%3.1 zero-effort integrable (build system-related files, no include-boilerplate or preprocessor defines)
%3.2 quality of service (powered by unit tests and benchmarks)

As previously described, we designed our generator pipeline to be extendable, allowing additional GPOs to be appended after the core pipeline. 
We implemented two GPOs, yet there are various conceivable extensions. 
The first one is designed to ensure the quality of service concerning the generated library; the latter increases the usability. 
We explain both GPOs in the following, closing the section with an outlook into further imaginable extensions.

\subsection{\textbf{Test Generation}}
\label{subsec:test_generation}

Every abstraction library requires a substantial demand for reliability. 
Hence, the provided functionality has to be tested extensively. 
In hand-written SIMD abstraction library solutions, test code typically resides beside the actual library code. 
The main obstacle here is that the library consists of two levels of detail. 
On the one hand, there are functions with expected behavior. 
On the other hand, there are hardware-dependent implementations for this very function. 
Typically, one or many test cases are implemented for one function, and every test case has to be carried out for each implementation. 
Testing all implementations requires either executing the tests on each supported hardware or using hardware-vendor-provided emulators. 
In addition, as we presented earlier, there can be multiple implementations for the same hardware, depending on specific instruction set extensions. 
Those implementations need specific compiler passes to be activated or deactivated. 
Another crucial concern for low-level testing code is that the tests often rely on the existing and proper functioning of other low-level code. 
For example, we have to transfer data into a SIMD mask type to test the primitive presented in Figure~\ref{lst:primitive} (b). 
While this may be a scalar value and thus can be assumed to work correctly, Intel's SSE, AVX, and ARM's Neon use SIMD registers as mask types. 
Consequently, transferring data from memory into a SIMD register must be reliable. 
This implies the necessity of a load primitive which again has to be tested before the primitive is used in another test case. 
To sum up, holistic testing of a hardware abstraction library requires a relatively complex testing infrastructure consisting of dependency checking, multiple build phases utilizing different \acp{sise}, and execution phases running on different machines or within emulators. 
Furthermore, every change within the library must also be reflected within this infrastructure. 

\begin{figure}
    \begin{minipage}[t]{.49\linewidth}
        \centering 
\begin{minted}{cmake}
cmake_minimum_required(VERSION 3.13)
set(Python_FIND_VIRTUALENV FIRST)
find_package(Python3 REQUIRED)
function(create_tsl)
  # parse parameters like:
  #  GENERATOR_ROOT, GENERATOR_DEST
  execute_process(
    COMMAND "${Python3_EXECUTABLE}" 
      -c "import cpuinfo; print(*cpuinfo.get_cpu_info()['flags'])"
    OUTPUT_STRIP_TRAILING_WHITESPACE OUTPUT_VARIABLE TARGETS_FLAGS
  )
  execute_process(
    COMMAND "${Python3_EXECUTABLE}" 
      "${GENERATOR_ROOT}/main.py" 
      -o "${GENERATOR_DEST}" 
      --targets "${TARGETS_FLAGS}"
    ANY
  )
  set(TSL_INCLUDE "${GENERATOR_DEST}/include" CACHE STRING)
\end{minted}
        \small (a) Static library top-level CMakeLists.txt.\\
    \end{minipage}
    \hfill
    \begin{minipage}[t]{.49\linewidth}
        \centering 
        %\lstinputlisting[frame=off,language=C++]{#4}
\begin{minted}{cmake}
cmake_minimum_required(VERSION 3.13)
add_library(tsl INTERFACE)
target_sources(simdlib
  INTERFACE include/tsl.hpp   
  # all generated header files
)
target_compile_options(
  tsl INTERFACE <TARGET_FLAGS>)
\end{minted}
        \small (b) Excerpt of the generated CMakeLists.txt.\\
        \centering 
        %\lstinputlisting[frame=off,language=C++]{#4}
\begin{minted}{cmake}
include(<path to generator>/tsl.cmake)
create_tsl(
  GENERATOR_ROOT "<GENERATOR_ROOT>"
  GENERATOR_DEST "${GENERATOR_ROOT}/tsl"
)

add_subdirectory(${CMAKE_CURRENT_BINARY_DIR}/tsl)
target_include_directories(<target> TSL_INCLUDE)
target_link_libraries(<target> tsl)
\end{minted}
        \small (c) Required changes to existing CMake projects to use our generated SIMD abstraction library.\\
    \end{minipage}
    \caption{Integration of library generation through cmake.}
    \label{lst:cmakelists}
\end{figure}

In our framework, we aim to reduce complexity for the library developer. 
Therefore, we decided to require test cases to be located within the primitive data UDP files preceding the actual implementations. 
For every primitive, multiple tests can be defined. 
Providing a test is optional, however, if one is specified, it consists of a mandatory implementation and a name for identification and it may have a field specifying the required primitives (dependencies) to ensure the integrity of the test. 
Our test-generation GPO works as follows. 
First, all primitives are checked for existing test cases. 
If no test cases are defined, a warning will be emitted. 
Then, a directed acyclic dependency graph is generated from all found test cases for the hardware-dependent library, where a node represents a concrete test case for an \ac{sru} and a data type, and an edge is created between node $a$ and $b$ if $b$ depends on $a$. 
If a test requires an untested concrete primitive, the test is marked as unsafe and will generate a warning message when the test is executed. 
%After the graph is built, we sort the nodes topologically so that every test is executed after all required precondition tests are passed. 
After the graph is built, we sort the nodes topologically to ensure, that a test is always executed after all of its required precondition tests are passed.

The test cases are embedded into the C++-testing framework Catch2~\cite{catch2}, which offers a single-header integration and consequently reduces the dependencies of our framework. 
To further reduce redundancy and complexity, we provide powerful helper functions for setting up the test and tearing it down again, including but being not limited to memory acquisition, comparing, and deallocation.

As our framework generates a hardware-dependent library, the test cases are also hardware-dependent. 
However, as described in the previous section, we can pass arbitrary hardware specifications into the generator pipeline, tricking the generator into assuming specific hardware as given. 
Consequently, running tests for different hardware merely requires a repeated invocation of our generator framework with different hardware specification input and execution of those tests on the target hardware or within an emulator. 
We consider the automatic invocation of emulators to be a prime candidate for further work. 

\subsection{\textbf{Build Environment Files Generation}}
\label{subsec:build_file_generation}
An important challenge we observed with existing hand-crafted libraries is their integration into existing systems. 
On the one hand, they typically enforce multiple libraries to be present on the system as dependencies. 
On the other hand, they require manual integration by setting up compiler flags, including the necessary files, and introducing substantial boilerplates for using the functionality. 
However, our framework relies on CMake as a de-facto-standard build-environment and aims at zero-effort integrability~\cite{DBLP:conf/icse/NguyenNP22}.
Thus, we developed a GPO that generates a CMake file containing all relevant header and source files and defines an interface library from those files (see Figure~\ref{lst:cmakelists} (b)). 
If the library should be created before building the dependent target, we added a static file, that determines the hardware flags and executes the generator (see Figure~\ref{lst:cmakelists} (a)) directly from within the cmake workflow. 
The generated library can be used by any project by just linking it against the target. 
Thus, for the build environment, all it takes to use the generated library from an existing project is to change the top-level \emph{CMakeLists.txt} and the \emph{CMakeLists.txt} where the targets are specified (see Figure~\ref{lst:cmakelists} (c)). 
The top-level header of our generated library must be included in the code, and the library-specific namespace must be introduced. 
Thereby, we substantially reduced the overhead for including and using our generated library. 

\subsubsection*{\textbf{Further Extensions}}
\label{subsec:further_extensions}
We already presented two valuable extension to our \emph{TSLGen} core pipeline, however, various further imaginable extensions exist. 
Prime candidates may be benchmark and documentation generation. 
For example, benchmarking is a crucial part of ensuring the quality of service. 
This especially holds in the context of hardware-near abstraction libraries. 
For a given primitive in a hardware-oblivious SIMD library, multiple implementations (variants) may exist. 
Those variants will produce the same result for a given input, yet they may perform very differently. 
To reason about which variant is superior to all others is hard to determine and error-prone, especially if the underlying hardware changes. 
Thus, we recommend benchmarking all variants within the generation process and choosing the best-performing one. 
Generally, executing the same code on different hardware platforms is almost guaranteed to yield different performance characteristics.
Further, due to wear-out effects of the hardware itself, an algorithm may behave differently over time, even when executed on the same machine.
Following this observation, a previously underperforming implementation could now be the best.
%Furthermore, in the context of writing data parallel code for database engines, we observed significant discrepancies in the execution time when running the same code on the one hand on different hardware and the other hand on the same hardware, but at moments that are quite different and far apart in time. 
Consequently, we argue that benchmarking alongside adaptive variant selection should be integrated as an ongoing process to achieve sustained performance over time ensuring the quality of service. 

\subsection{Toolsupport}
\label{subsec:tooling}
Due to the decision to separate templates and data models, it is only necessary to adapt the data model to extend the functionality of the generated library. 
While any \mbox{(semi-)structured} format is sufficient for the data model, we have opted for YAML due to its compactness and implicit extensibility. 
However, this extensibility and flexibility can also lead to difficulties, for example, with the input and type safety validation. 
Furthermore, the debugging of YAML files can be cumbersome, especially if users employ either tabs or multiple spaces as indentations. 
To meet these challenges, we provide a VSCode extension that supports auto-completion, real-time rendering, and indexing of the data model, among other things. 
%The auto-completion could be realized surprisingly straightforwardly as we conceived a schema for the data model with mandatory and optional fields and types. 
The auto-completion could be realized surprisingly straightforwardly as our data model already provides the mandatory and optional fields and types.

\section{Use-Case Studies}
\label{sec:Evaluation}
% \usepackage{multirow}

%\begin{figure*}
%\begin{tabularx}{\textwidth}{lc@{\extracolsep{\fill}}r}
%\parbox[t][][t]{0.95\columnwidth}{\inputminted{yaml}{supplementary/hadd_sse.yaml}} & \hspace*{\fill} &  %\multirow[c]{2}{*}{\parbox[t][][t]{0.95\columnwidth}{\inputminted{yaml}{supplementary/hadd_fpga.yaml}}} \\
%\parbox[t][][t]{0.95\columnwidth}{\inputminted{yaml}{supplementary/hadd_arm.yaml}} & \hspace*{\fill} &                   
%\end{tabularx}
%\caption{User provided data for the primitive \texttt{hadd} for Intel SSE, ARM Neon and Stratix 10 FPGA. While the definition for Neon and the FPGA %are the same for all supported arithmetic types, SSE needs specific treatment per type class. For reasons of limited space we only present the %definition for (unsigned) integer values.}
%\label{lst:hadd}
%\end{figure*}

%    \hfill
%    \begin{minipage}[t]{.47\linewidth}
%        \centering \textbf{(b) Implementation for ARM Neon}\\
%        \inputminted{cpp}{supplementary/algo_tsl.cpp}
%    \end{minipage}
%    \caption{User provided data for the primitive \texttt{hadd} for Intel SSE, ARM Neon.}
%    \label{lst:hadd}
%\end{figure*},linenos=no
\begin{figure*}
    \begin{minipage}[t]{.49\linewidth}
\begin{minted}{cpp}
#include <hwy/aligned_allocator.h>
#include <hwy/highway.h>
/* ... */
HWY_BEFORE_NAMESPACE();  
namespace HWY_NAMESPACE {   namespace hn = hwy::HWY_NAMESPACE;
uint32_t filter_count(float const * data, size_t count, float l, float u) {
  const hn::FixedTag<float, HWY_LANES(float)> simdT;
  const hn::FixedTag<uint32_t, HWY_LANES(uint32_t)> simdTui32;
  auto rb = hwy::AllocateAligned<uint32_t>(Lanes(simdT));
  auto rv = Zero(simdTui32);
  const auto iv = Set(simdTui32, 1);
  const auto lv = Set(simdT, l);
  const auto uv = Set(simdT, u);
  for(size_t i = 0; i < count; i += Lanes(simdT)) {
    const auto dv = Load(simdT, &data[i]);
    const auto cv = And((dv >= lv), (uv >= dv));
    const auto irv = 
      VecFromMask(simdTui32, RebindMask(simdTui32,cv)) & iv;
    rv += irv;
  }
  Store(rv, simdTui32, rb.get());
  auto r = rb[0];
  for(size_t i=1; i < Lanes(simdTui32); ++i) {
    r += rb[i];
  }
  return r;
}}
HWY_AFTER_NAMESPACE();
\end{minted}
        \centering \small (a) Implementation using Google's HWY
    \end{minipage}
    \hfill
    \begin{minipage}[t]{.49\linewidth}
\begin{minted}{cpp}
#include <tslintrin.hpp>

/* ... */
template<typename SRU>
uint32_t filter_count(float const * data, size_t count, float l, float u) {
  using namespace tsl;
  using CSRU = simd<uint32_t, typename SRU::target_extension>;
  
  auto rv = set1<CSRU>(0);
  const auto iv = set1<CSRU>(1);
  const auto lv = set1<SRU>(l);
  const auto uv = set1<SRU>(u);
  for(size_t i = 0; i < count; i += SRU::vector_element_count()) {
    const auto dv = load<SRU>(&data[i]);
    const auto cv = between_inclusive<SRU>(dv, lv, uv);
    const auto irv = 
      binary_and<CSRU>(reinterpret<SRU, CSRU>(to_vector<SRU>(irv)), iv);
    rv = add<CSRU>(rv, irv);
  }

  return hadd<CSRU>(rv);
}
   
\end{minted}
        \centering \small (b) Implementation using our generated TSL
    \end{minipage}
    \caption{Possible implementation of an algorithm counting the number of elements in a given range.}
    \label{lst:algo_hwy}
\end{figure*}

For assessing our \emph{TSLGen}, we present two case studies, first an algorithm that stems from the field of database systems and second a single primitive. 
In the following, we will describe both case studies and evaluate our generator approach concerning applicability and extensibility.

\subsection{Case Studies Description}

%%%%%% Range Count %%%%%%
As our first evaluation case study, we choose an algorithm that counts the occurrence of elements within a given range from an input. 
The input is usually a large array of arithmetic values (e.g., integer or floating point). 
Such an algorithm can be seen frequently within database and machine learning systems. 
We present a possible implementation of such an algorithm using a SIMD abstraction library (i) developed by the industry (see Figure~\ref{lst:algo_hwy}(a)) and (ii) generated from our TSLGen generator framework (Figure~\ref{lst:algo_hwy}(b)). 
The algorithm can be subdivided into three phases: (i) initialization, (ii) processing, and (iii) finalization. 
For the \emph{initialization phase}, the result SIMD register (\emph{rb}), the increment SIMD register (\emph{iv}), and the range-limits SIMD registers (\emph{lv} and \emph{uv}) are set up by broadcasting a single value to all lanes of the corresponding SIMD register (see lines 10-13). 
The \emph{processing phase} (lines 14-20) mainly consists of three steps. 
First, $N$ consecutive elements are transferred into the SIMD register \emph{dv}, while $N$ equals the number of available lanes within a SIMD register. 
Assuming that \emph{float} values have a size of 4-byte that leads to $N=4$ for 128-bit wide, $N=8$ for 256-bit wide and $N=16$ for 512-bit wide SIMD registers, respectively. 
Subsequently, the values within \emph{dv} are compared against the given range. 
The result of the comparison is stored in \emph{cv}. 
Please note that the result of a comparison intrinsic for Intel's SSE, AVX, and ARM's Neon results in a SIMD register, where all bits in a lane are set to 1 if the comparison evaluates to true, 0 otherwise. 
In contrast, AVX512 produces a scalar value, where the $n$-th bit is set to 1 if the compared value meets the condition and 0 otherwise. 
Thus, the type of \emph{cv} is not determined for the given code and will become deduced at compile time. 
To circumvent this issue, we create a SIMD register from the mask, which results in a masked broadcast for AVX512 and a \emph{NOP} for all remaining SISEs. 
The resulting SIMD register is binary ANDed with the \emph{iv}, and the result is stored in the SIMD register \emph{irv}. 
Consequently, if the $n$-th element in \emph{dv} was in the requested range, the $n$-th element of \emph{irv} will contain a 1, 0 otherwise. 
This is then added to and stored in the result SIMD register. 
While up to this point, the implementations using HWY and TSL are mostly similar, the last (\emph{finalization}) phase ((a): lines 21-25; (b): line 21) differs between both variants since TSL provides a function to add up all values from a given SIMD register. 
In contrast, when using HWY, we need to store the SIMD register in a buffer and aggregate the values in a scalar manner\footnote{We also implemented another flavor of this algorithm (SELCOUNT\_popcnt), omitting the result register by carrying out a population count of the resulting mask created by using the primitive depicted in Figure~\ref{lst:msb_extract_ghighway} and adding the number of set bits directly to the result.}. 

\begin{figure}
    \begin{minipage}[t]{\linewidth}
\begin{minted}{yaml}
- target_extension: "sse"
  ctype: ["uint32_t", "int32_t"]
  lscpu_flags: ["sse2", "ssse3"]
  specialization_comment: "Signed Addition."
  is_native: False
  implementation: |
    auto const res = _mm_hadd_epi32(value, value);
    return _mm_cvtsi128_si32(res) + _mm_cvtsi128_si32(_mm_bsrli_si128(res, sizeof(uint32_t)));
- target_extension: "neon"
  ctype: 
    ["uint8_t", "int8_t", "uint16_t", "int16_t", "uint32_t", "int32_t", "uint64_t", "int64_t", "float", "double"]
  lscpu_flags: [ 'neon' ]
  implementation: "return vaddvq_{{ extension_dependent_suffix[ctype] }}(value);"
\end{minted}
    \end{minipage}
    \caption{User provided data for the primitive \texttt{hadd} for Intel SSE, ARM Neon.}
    \label{lst:hadd}
    %\vspace{-0.4cm}
\end{figure}

%%%%%% HADD %%%%%%
As our second case study, we choose the horizontal aggregation primitive (\texttt{hadd}) (see Figure~\ref{lst:hadd}). 
The primitive expects a SIMD register (value), accumulates all values within the register, and returns the sum. 
Intel SSE does not provide such functionality natively. 
Consequently, the corresponding definition consists of multiple steps. 
First, adjacent pairs of 32-bit entries are added so that \emph{res} contains the sum of the lower two values from our input register in the lowest 32-bit and the sum of the upper two values in the second 32-bit, respectively (see line 7, target extension: "sse"). 
For the final result, the lower 32-bits are extracted (see line 8) and added to the next 32-bits (by shifting the content of the register by 32 to the right and extracting the lower 32-bits again). 
The corresponding definition includes even more steps for smaller value types like 16-bit or even 8-bit integers. 

In contrast to SSE, ARM Neon natively supports the needed horizontal aggregation for all relevant arithmetic types. 
The intrinsics from Neon follow a very strict naming schema, including input and result data types. 
The input type is encoded as a combination of a single character for the type and the size in bits (e.g., u8 for \textbf{u}nsigned \textbf{8}-bit values). 
Our generator approach allows implementations to contain templates, which will be rendered for every given type individually. 
Within the user-provided data of the Neon extension, a dictionary is defined, which contains the described mapping. 
Consequently, the implementation of all implementations for Neon consists of a single line of code. 
%The FPGA extension uses a nested loop which will be unrolled by the compiler and forms an adder tree. 
%We will describe the implementation in section \ref{subsec:extensibility}.

\subsection{Applicability}
\label{subsec:applicability}

To evaluate the applicability of our proposed generator framework \emph{TSLGen}, we implemented the range-count algorithm from Figure~\ref{lst:algo_hwy} using Google Highway (HWY)~\cite{highway} and our generated TSL library. 
As shown in Figure~\ref{lst:algo_hwy}, the two implementations are similar. 
They only differ in the primitive names and a fundamental design decision, where HWY uses tag-dispatching to decide which primitive implementation should be used while TSL uses template parameters. 
Both approaches will result in the same binary code most likely. 
We build the code using clang++-10 with optimization flag -O2 on a Xeon Gold 6240R. This processor supports all recent Intel SISEs up to AVX512(F, DQ, BW, VL, VNNI). 

To evaluate the performance impact of both libraries, we conducted a micro-benchmark, where the implemented algorithms processed $4GiB$ of integer data using either 128-, 256- or 512-bit wide SIMD registers with the associated primitives. 
The data was generated randomly using a uniform distribution in the $[0,100000]$ range, and the lower and upper predicate values were set to 5 and 15, respectively. 
Since the algorithm is branch-free, the data and the range do not contribute to the performance since the number of executed instructions merely depended on the input size. 

\begin{figure}
    \includegraphics[width=0.9\columnwidth]{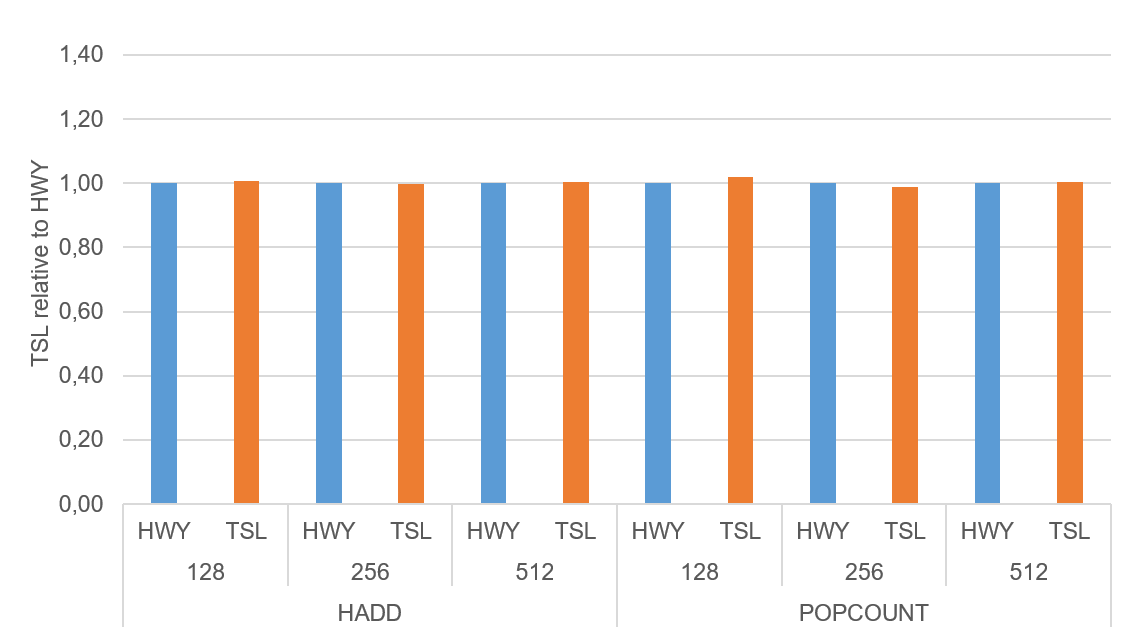}
    \caption{Relative runtime comparison for our case study (range count) using either a horizontal addition as depicted in \ref{lst:algo_hwy} (left) or a bit-population count.}
    \label{fig:bench_hwy_tsl}
\end{figure}

We ran our micro-benchmark 10 times and averaged the execution time. 
As we first want to check whether our generated library is comparable to an industry-forged one, we report on the relative differences between both in  Figure~\ref{fig:bench_hwy_tsl}. 
As this figure implies, both variants perform equally with a maximum deviation of $[-0.3\%,+0.6\%]$ in the runtimes. 
The other variant using a \emph{popcount} instead of a SIMD register \emph{ADD} produces similar results with a deviation of $[-1\%, +1.8\%]$. 

We also investigated further case studies leading to the same conclusion. 
The usage of our generated library only differs marginally from an application programming perspective, and the achieved runtimes are very similar. 
Thus, our generated library is equivalent to a hand-crafted one. 

\subsection{Extensibility}
\label{subsec:extensibility}
Intel's OneAPI allows C++ developers to write code that can be synthesized using the \emph{dpcpp} compiler as an FPGA image and run on such hardware. 
In combination with unified shared memory (USM), allowing the FPGA to directly access host memory, this opens up the opportunity for offloading various algorithms to FPGAs straightforwardly. 
To illustrate the extensibility of our approach, we defined an SRU for an FPGA within our user-provided data. 
Since FPGAs do not have predefined SIMD register types, we used fixed-sized arrays. 
Through our design decision of using class templates for representing SRUs that have, besides others, a register size as a non-type template parameter, we can define a SIMD-length agnostic SRU like it is proposed by ARM SVE. 
However, to support arbitrary SIMD register lengths, we had to change the primitive templates since, for all prior considered SISEs, the register size is predefined. 
In total, we had to add 19 lines of code to our schema and templates before integrating FPGA into our generator. 

\begin{figure}
    \begin{minipage}[t]{\linewidth}
\begin{minted}{cpp}
template<typename T, int VS>
using fpga_register_t [[intel::fpga_register]] = std::array<T, (VS>>3)/sizeof(T)>;

template<typename T, int VS> [[nodiscard]] T hadd(fpga_register_t<T,VS> const values) {
  fpga_register_t<T,S> result;
  auto pos = 0;
  #pragma unroll
  for (; pos < (result.size()>>1); ++pos) {
    result[pos] = values[pos<<1] + values[(pos<<1)+1];
  }
  int base = 0;
  #pragma unroll 
  [[intel::loop_coalesce((VS>>4)/sizeof(T)), intel::ivdep(result)]] 
  for (size_t stage = (result.size()>>2); stage >= 1; stage>>=1) {
    for (size_t i = 0; i < stage; ++i) {
      result[pos++] = result[base + (i<<1)] + result[base + (i<<1)+1];
    }
    base += stage<<1;
  }
  return result[(VS>>3)/sizeof(T)-2];
}
\end{minted}
    \end{minipage}
    \caption{Implementation of a horizontal add which compiles and runs on an Intel Stratix 10. This implementation can be used to add a new primitive definition straightforwardly.}
    \label{lst:hadd_fpga}
\end{figure}

As a proof of concept, we added implementations for all necessary primitives from our range-count use case (see Figure~\ref{lst:algo_hwy}(b)). 
We implemented most primitives using a loop over the register type (aka. array) and executing the specific instruction on every element. 
However, this schema is insufficient for horizontal operations like the \emph{hadd} primitive from our use-case (see Figure~\ref{lst:hadd_fpga}). 
Such operations are typically realized using an adder tree where elements are added pairwise, and the results are again added pairwise until a single value is produced. 
The depth of such an adder tree equals $log_2(N)$, where N is the number of elements. 
Consequently, if N is known at implementation time, the algorithm executes a fixed number of instructions. 
However, since our FPGA SRU does not specify a SIMD-register size and $N = \frac{size}{sizeof(type)}$, we had to come up with a size and type agnostic algorithm. 
Our implementation (see Figure~\ref{lst:hadd_fpga}) consists of a nested loop, where the outer loop determines the number of executions, that will be carried out in the inner loop, starting with half of the element count and halving on each iteration. 
The inner loop adds adjacent values within the array and consecutively stores the result in the same array. 
For example, assuming a 16-element wide register, the outer loop would be executed four times. 
Eight adjacent pairs are added in the first iteration, and the results are stored in the lower half of the register. 
In the second iteration, the lower eight adjacent pairs are added and stored in the lower quarter of the register, and so on. 
As shown in lines 7 and 12, we use compiler pragmas to instruct the compiler to unroll the loops accordingly. 
Thus, the compiler will generate an adder tree as described previously. 

\begin{figure}
    \includegraphics[width=0.9\columnwidth]{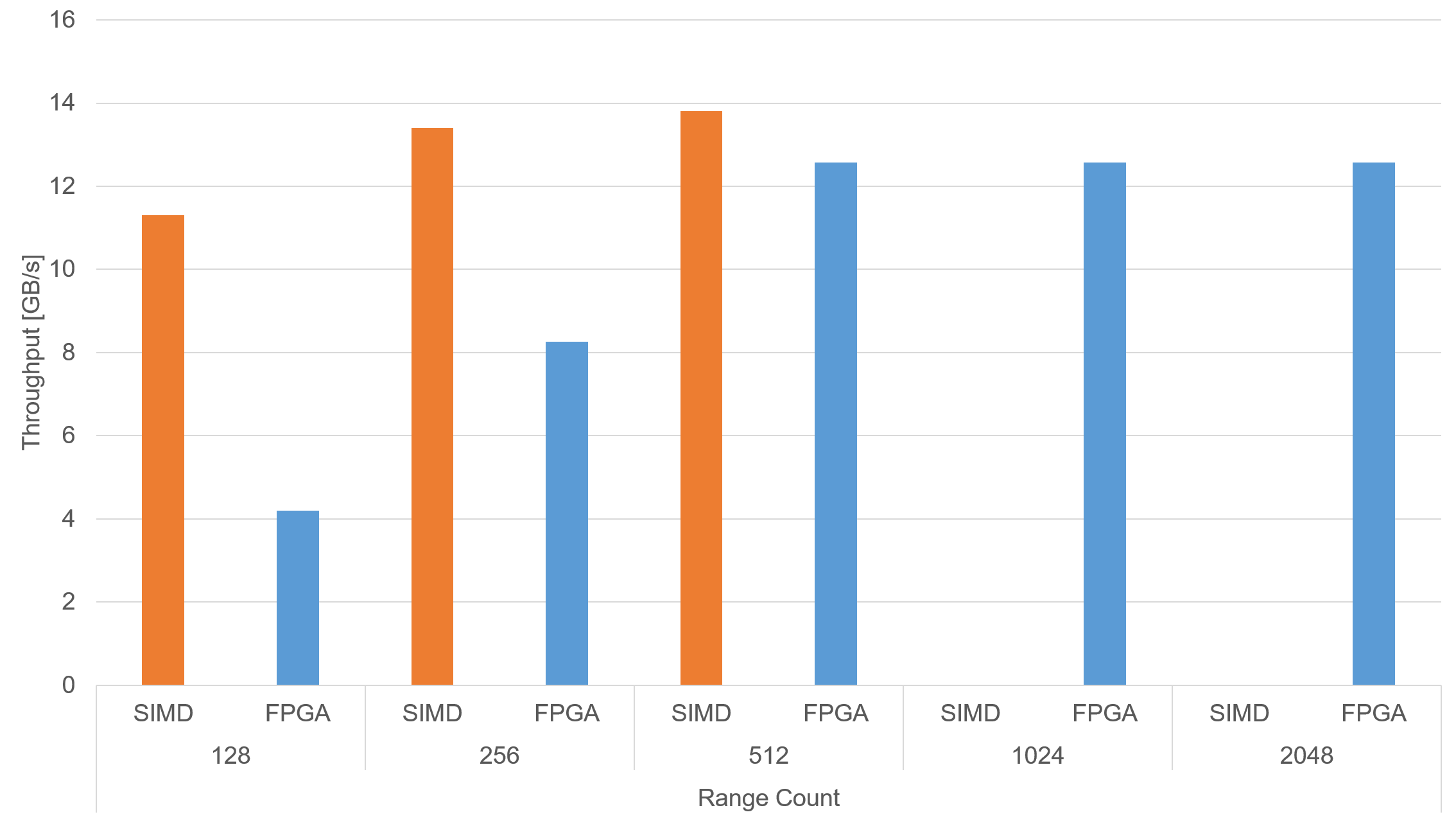}
    \caption{Throughput of our case study (range count) using a horizontal addition as depicted in Figure~\ref{lst:algo_hwy} on 4GiB data running using either Intel SIMD SISEs or FPGA (on Stratix 10).}
    \label{fig:bench_simd_vs_fpga}
    %\vspace{-0.4cm}
\end{figure}

In total, the seven necessary primitives required 100 lines of code, whereas, in contrast, the generated library code for supporting FPGA comprises 3581 lines. 
After the integration, we could use the very same implementation of our use-case algorithm and call it from within a lambda function which is passed to \emph{sycl::handler} as \emph{single\_task} and added to a \emph{sycl::queue}. 
We build our micro-benchmark from Section~\ref{subsec:applicability} again on Intel's DevCloud using dpcpp\footnote{dpcpp version 2022.2.0.20220730}. 
We repeated our experiment, where the SSE, AVX, and AVX512 variants ran on an Intel Xeon Platinum 8256, and the FPGA code was offloaded to and then executed on an Intel Stratix 10. 
Since the used FPGA supports unified shared memory (USM), the FPGA kernel can directly read from the host memory and write the result back. 
Consequently, the throughput of the FPGA execution is limited by the PCIe bandwidth if the executed algorithm is pipelined. 

We ran our micro-benchmark on the FPGA using five different vector sizes from 128-bit up to 2048-bit. 
As depicted in Figure~\ref{fig:bench_simd_vs_fpga}, the variant using 512-bit wide registers achieves peak performance close to 12 GiB/s throughput and thus mostly saturates the bandwidth. 
That is because the Stratix 10 is limited to 512b/s per channel. 
The available processing units are only partially utilized using smaller or bigger register sizes. 
In total, it took us approximately one hour to integrate the necessary functionality for running our micro-benchmark on the Stratix 10. 
However, it should be noted that the relevant instructions were comparable and straightforward to conceive. 
Nevertheless, we argue that integrating completely new hardware, which is not even designed as a SISE, powerfully demonstrates the extensibility and advantages of our approach.

\section{Discussion}
\label{sec:Discussion}
%GPU passt hier irgendwie nicht rein
Now that we have generally introduced and evaluated our generator framework \emph{TSLGen}, the question is, "What is it good for"? 
Code generation, as such, is not new but a widely used technique. 
Apart from the apparent advantages already described, separating structure-relevant information from the actual data model offers at least two exciting advantages in our particular use case. 

Firstly, in contrast to a holistic hand-crafted library, with our data model, we have a directed mapping of function names to their hardware-dependent implementation. 
While it can drastically reduce redundancy and thus improve readability by its very nature, the data model and its implicit mapping enable structural reasoning regarding complex and usually hidden dependencies on the one hand. 
On the other hand, the mapping can be inverted, which has the potential to enable a (semi-)automatic generalization of existing explicitly simplified code. 
The basic idea is to identify functionalities from the data model in any code base and replace them with the corresponding library function calls. 
While this can be easily realized for trivial functions by string matching, abstract rules probably have to be derived for more complex functions to enable recognition and substitution. 
Consequently, explicitly simdified existing code could be generalized with minimal effort and thus, as already shown, be executed not only on different SIMD extensions but even on accelerators such as FPGAs. 
This would pave the way for hardware portability of existing code without substantial manual effort, which becomes increasingly necessary, considering the ongoing trend in hardware heterogeneity and specialization. 

Another advantage of separation is the possibility of generalizing the data model to support different languages. 
Thus, using the existing function implementations to generate a RUST (or other SIMD supporting language) library may be possible. 
Since RUST provides SIMD intrinsics similar to C/C++ and is also comparable in its type system, we argue that only an adaptation of the structure templates is necessary to obtain a SIMD hardware abstraction library for RUST.
Thus, both discussed aspects should be investigated in future work and our developed framework \emph{TSLGen} would provide the ideal basis for this.

\section{Related Work}
\label{sec:RelatedWork}

Generally, the relationship between hardware and software is at an inflection point. 
While software benefited from higher clock cycles of modern CPUs for many years, hardware components are advancing at an incredible speed, providing a rich bouquet of novel techniques. 
For example, in the area of computing elements, the core count increased and internal techniques like advanced SIMD instruction set extensions, pre-fetching, or branch prediction dramatically improved within modern CPUs. 
Moreover, alternative computing element approaches like GPUs or FPGAs have been developed.
With these new and increasing opportunities that open up the emerging hardware landscape, there also arises a plethora of new challenges regarding application implementation on these emerging hardware platforms. 
To solve these new programming challenges, state-of-the-art approaches are based on general-purpose or domain-specific frameworks.

\subsubsection*{\textbf{General-Purpose Programming Frameworks}}

One approach for implementing applications on different hardware platforms are general heterogeneous programming frameworks, such as OpenMP~\cite{dagum1998openmp}, OpenCL~\cite{munshi2011opencl}, CUDA~\cite{sanders2010cuda}, SYCL~\cite{DBLP:conf/parco/ReyesL15}, and oneAPI~\cite{reinders2021data}. 
These models are not specialized for a single domain or class of applications but define abstraction mechanisms and language extensions for common computational patterns, such as parallel loops or concurrently executing tasks. 
Moreover, they are tightly integrated with an existing serial programming language, such as C, C++, or Fortran, and thus allow at least partial re-use of an existing serial code base.

The advantage of the generic nature of these programming models is that they can be used to accelerate a wide range of applications from very different domains, as they focus on fundamental computational patterns rather than domain-specific abstractions.
Improvements in implementing a programming model, e.g., in the compiler or runtime library, benefit many applications, and hardware vendors only need to implement the abstractions of the programming model once to open up their hardware to many users and applications.
On the other hand, due to their general design, such programming frameworks fail to capture the domain-specific, high-level semantics of hardware platforms and still operate on a comparably low level of abstraction, e.g., reasoning about individual loops. 
For example, SIMD often plays no or only a subordinate role~\cite{DBLP:conf/iwomp/KlemmDTSCM12}.
Memory management is typically also explicit and requires careful attention from the programmer. 
Because the language mechanisms (e.g., pragmas) are directly integrated into program code, applications implemented in such general frameworks also mix what is computed with how it is computed, with adverse effects on code readability and maintainability.

As the abstractions of general programming frameworks still operate on a comparably low level, implementations using such a framework are often specific to a single class of devices. 
One example of this problem can be found when comparing vendor recommendations for programming FPGAs or GPUs using OpenCL: While FPGA vendors usually suggest a single work-item implementation, with FPGA-specific optimizations such as pipelining applied inside that single work-item, GPU vendors usually suggest parallelizing an application across many work-items and, if applicable, also concurrently active kernels. Consequently, OpenCL implementations optimized for FPGAs usually do not deliver good performance on GPUs and vice versa.

\subsubsection*{\textbf{Domain-Specific Programming Frameworks}}
In contrast, domain-specific programming frameworks can be used to overcome the limitations of general-purpose programming models.
They specialize in a single domain and provide programming language abstractions for operations specific to this domain. 
Because these domain-specific operations have well-defined semantics and the memory access pattern is known beforehand, it is possible to separate what is computed from how it is computed by moving the mapping of domain-specific operations to hardware platforms facilities into the implementation of the programming framework. This also allows the mapping to vary depending on available hardware and the desired performance independently of the actual source code. 
The SIMD abstraction libraries are a prominent example~\cite{DBLP:conf/IEEEpact/EsterieGFLR12,highway,TVLPaper,xsimd,DBLP:conf/ppopp/WangWTSM14} as described in Section~\ref{sec:Background}. 
Other similar SIMD-related approaches are Weld~\cite{DBLP:journals/pvldb/PalkarTNTPNSSPA18}, Voodoo~\cite{DBLP:journals/pvldb/PirkMZM16}, or Sierra~\cite{DBLP:conf/ppopp/LeissaHH14}. 
The downside of domain-specific frameworks is the additional effort that initially must be spent to define and implement the framework itself. 
Moreover, maintenance and extensibility are critical points. 
As presented in this paper, our developed framework \emph{TSLGen} tackles these issues using code-generation features for SIMD abstraction libraries. 

\subsubsection*{\textbf{Auto-vectorization}}

Due to the complexity of explicit SIMD programming, there is also a long history of using compilers to automatically translate scalar code into vectorized code with SIMD instructions for different hardware platforms~\cite{DBLP:conf/nips/MendisYPAC19,DBLP:conf/cgo/NuzmanH06,DBLP:conf/IEEEpact/NuzmanZ08}.  
Nowadays, all major commercial and open-source compilers have auto-vectorization capability. 
However, the evaluation in~\cite{DBLP:conf/IEEEpact/MalekiGGWP11} showed that compilers automatically vectorize only a few loops from real applications.
In general, auto-vectorization is a mature technology working well on simple logic and regular loops with no unknown side effects.
Thus, explicit SIMD programming is still state-of-the-art to fully utilize the power of SIMD. 
Based on that, SIMD abstraction layers are the way to go to address the heterogeneity challenge.

\section{Conclusion}
\label{sec:Conclusion}

\emph{Single Instruction Multiple Data (SIMD)} is a heavily-used hardware-driven performance technique in various application domains.
Moreover, the SIMD extensions of modern CPUs are growing in SIMD register size and the complexity of the provided SIMD instruction sets. 
%Since using MIMD alongside and/or combined with SIMD are reliable techniques for optimizing algorithms, existing hardware constraints must be considered, especially concerning the used SIMD instructions and the utilized degree of data-level parallelism, to achieve optimal speedups. 
SIMD abstraction libraries have established themselves as de-facto standard to tackle that. 
Unfortunately, existing libraries are hand-written and inherently complex, making maintainability and extensibility difficult. 

Additionally, directly usable hardware tends to become more and more diverse while also being increasingly specialized at the same time, like GPUs and FPGAs. 
This development offers outstanding opportunities for compute-heavy and data-centric applications on the one hand. 
On the other hand, integrating new hardware into hand-written libraries may entail disruptive changes to the core design. 

To overcome that and to establish a future-proof SIMD abstraction library, we have presented \emph{TSLGen}, a novel framework approach for generating a \emph{Template SIMD Library (TSL)} in this paper. 
Our evaluation shows that the application programming effort is comparable to existing libraries, and we achieve the same performance results. 
However, our generator approach is easy to maintain and extend.

\end{document}